\documentclass[12pt]{iopart}
\usepackage{graphicx} 
\usepackage{color}
\usepackage{bm}

\begin{document}

\newcommand \be {\begin{equation}}
\newcommand \ee {\end{equation}}
\newcommand \bea {\begin{eqnarray}}
\newcommand \eea {\end{eqnarray}}

\title[Extreme statistics and volume fluctuations in a confined one-dimensional gas]{Extreme statistics and volume fluctuations in a confined one-dimensional gas}

\author{Eric Bertin$^1$, Maxime Clusel$^2$, Peter~CW Holdsworth$^1$}

\address{$^1$ Universit\'e de Lyon, Laboratoire de Physique,
\'Ecole normale sup\'erieure de Lyon, CNRS,
46 all\'ee d'Italie, F-69007 Lyon, France\\
$^2$ Department of Physics and Center for Soft Matter Research,
New York University,\\
4 Washington Place, New York, NY 10003, United States of America
}
\ead{eric.bertin@ens-lyon.fr, maxime.clusel@nyu.edu and peter.holdsworth@ens-lyon.fr}
\begin{abstract}
We consider the statistics of volume fluctuations in a
one-dimensional classical gas of non-interacting particles confined
by a piston, and subjected to an arbitrary external potential.
We show that despite the absence of interactions between particles,
volume fluctuations of the gas are non-Gaussian, and are described by
generalized extreme value distributions. The continuous shape parameter
of these distributions is related to the ratio between
the force acting on the piston, and the force acting
on the particles. Gaussian fluctuations are recovered
in the strong compression limit, when the effect of the external potential
becomes negligible. Consequences for the thermodynamics are also discussed.
\end{abstract}


\section{Introduction}

Extreme value distributions, describing the
fluctuations of the k$^\mathrm{th}$ largest value in a set of random variables
\cite{Gumbel,Galambos}, have recently been shown 
to play a role in physics well-beyond their standard area of application.
Such distributions are indeed found analytically in different problems where 
extreme values are not involved {\it a priori}, such as global measures
in $1/f$-noise \cite{Antal,Gyorgyi}, in interfaces in random environment
\cite{Fedorenko03}, in coupled quantum oscillators \cite{Wijland04},
or in the level density of Bose gases \cite{Comtet07} and related problems
of integer partitions \cite{Comtet07b}.
Moreover, distribution functions describing the fluctuations 
of disparate global measures in complex systems \cite{BHP} can be well
approximated by extreme value distributions \textit{generalized}
to real values of the parameter $k$ \cite{Bramwell,BramwellPRE,Portelli}.
Examples also include Burgers turbulence \cite{Noullez02},
models of heterogeneous glassy dynamics \cite{Chamon04,Jaubert07},
relaxing granular gases \cite{Brey05},
and very recently, order parameter fluctuations in a liquid crystal
close to a critical point \cite{Joubaud08}.
Generalized extreme value distributions were also found analytically
in a stochastic cascade model with dissipation \cite{Bertin05}.

A scenario has been proposed recently to explain the rather puzzling
emergence of extreme value distributions in contexts not clearly related
to extreme processes \cite{Bertin05,BC06}. One can actually reformulate the
original problem of extremes as a problem of sums of non-identically and
generically correlated random variables, leading to the same extreme value
distributions. Provided that the joint probability of the summed random
variables has a particular form, extreme values could result from random 
sum problems. For instance the appearance of a Gumbel distribution
in the $1/f$-noise model is understood in a simple way within this framework
\cite{Bertin05}.
Furthermore this mapping provides a new view on generalized extreme value 
distributions. It is indeed possible to generalize the equivalent sum
problem simply by extending the joint distribution to real values of the
index $k$: generalized extreme value distributions then find a natural
interpretation as limit distributions of sums of non-identically and
generically correlated random variables \cite{Bertin05,BC06},
with a particular form of joint probability (see Eq.~(\ref{dist}) below).
However, with the exception of uncorrelated but non-identically distributed
variables which further
generalizes the $1/f$-noise model by including a low frequency
cut-off \cite{BC06}, the aforementioned class of correlated random
variables looks rather formal and far from physical applications.

In this paper, we illustrate on a very simple statistical model,
namely a one-dimensional classical gas of independent particles,
how the above class of random variables can find a natural application
in a physical context.
This type of model has a long history and has proved very useful for the
illustration of statistical concepts
\cite{Frisch,Jepsen,Lebowitz,Piasecki,VandenBroeck,Bena}.
However, at variance with most previous studies on such a model,
we consider here
a gas confined both by a piston and by an external potential.
We find that although the system is composed of independent
and identical particles,
the distributions of volume fluctuations are non-Gaussian, and described
in the limit of a large number of particles by generalized extreme value
distributions, indicating  that correlations appear in the
system\footnote{Another reason for non-Gaussian fluctuations
would be that the variance diverges, leading to L\'evy-stable
laws \cite{Petrov}, but this is not the case here.}.
The nature of the correlations is however rather subtle: the
local volume between particles is correlated, but their positions are not.

Note that connections between the one-dimensional classical gas of particles
(or Jepsen gas) and extreme value statistics have already been reported
when the gas can freely expand, in the absence of piston and potential
\cite{Bena}.
However,  these results concern the velocity of the rightmost particle
rather than the volume of the gas, and are thus of a different nature.

\section{A simple model of a one-dimensional confined gas}

\begin{figure}[t]
\centering\includegraphics[height=5cm,clip]{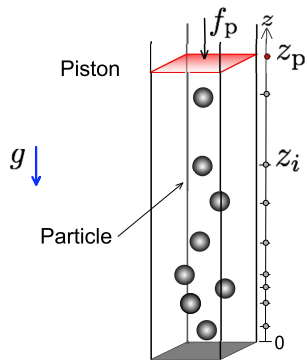}
\caption{\sl Schematic illustration of the model: point-particles at position
$z_i$ are placed in a container closed by a moving piston, at position
$z_\mathrm{p}$. An external potential (illustrated as the gravity $\bm{g}$)
acts on the particles, and a force $\bm{f}_\mathrm{p}$ (that may be of the same
origin as that acting on the particles) is applied on the piston.}
\label{schema}
\end{figure}

We consider an ideal system composed of $N$ point-particles
placed in a cylindrical container with long axis $z$ and with a small diameter
with respect to its length (quasi one-dimensional geometry).
The model is illustrated in a schematic way in Fig.~\ref{schema}.
The position of the particles along the $z$-axis is denoted as $z_i$,
$i=1,\ldots,N$.
A hard wall, acting as a reflecting boundary is placed at $z=0$,
so that particles are constrained to remain on the half-space $z>0$.
The container is closed by a piston that can freely move along the
$z$-axis. The position of the bottom of the piston is denoted $z_\mathrm{p}$,
so that the volume of the system is given by $V=S z_\mathrm{p}$, where
$S$ is the cross-section of the cylinder.
In addition, particles are subjected to an external potential $U(z)$, $z>0$,
that tends to confine them in the small $z$ region, that is $U(z)$
is assumed to be an increasing function of $z$.
The piston is also subjected to an external potential
$U_\mathrm{p}(z_\mathrm{p})$, that may
differ from the potential acting on the particle. The potential
$U_\mathrm{p}$ may for instance be the gravitational potential acting on
the piston. It may also be the potential caused by an operator exerting
a constant force $f_\mathrm{p}$ on the piston, corresponding to
a linear potential
$U_\mathrm{p}(z_\mathrm{p})=-f_\mathrm{p} z_\mathrm{p}$,
or by a spring fixed on the piston, corresponding to a quadratic potential
$U_\mathrm{p}(z_\mathrm{p})=\frac{1}{2}K(z_\mathrm{p}-z_0)^2$.
$U_\mathrm{p}(z_\mathrm{p})$ may also result from the superposition
of the different types of potentials mentioned above.

The container plays the role of a heat bath that thermalizes
the particles at a given temperature $T$.
Collisions between the particles
and the piston are elastic, so that the piston in turn thermalizes at
temperature $T$ (the piston does not have any internal structure,
only its translation degree of freedom thermalizes).
Note that the present model differs from the so-called Jepsen
gas \cite{Frisch,Jepsen,Lebowitz,Piasecki,VandenBroeck},
due to the presence of a heat reservoir and of an external potential.

\section{Distribution of volume fluctuations}
\label{sect-dist}

We now characterize quantitatively the volume fluctuations of
the system through the position, $z_\mathrm{p}$, of the piston.
When all the particles and the piston are equilibrated at temperature $T$,
the equilibrium distribution reads, with $z_i, z_\mathrm{p}>0$,
\be \label{eq-dist}
P_N(z_1,\ldots,z_N,z_\mathrm{p}) = \frac{1}{Z}\, e^{-\beta U_\mathrm{p}(z_\mathrm{p})}
\prod_{i=1}^N e^{-\beta U(z_i)} \, \Theta(z_\mathrm{p}-z_i),
\ee
where $\beta=1/k_B T$ is the inverse temperature, $Z$ is the partition function
and $\Theta(x)$ is the Heaviside function.
The volume distribution $Q_N(z_\mathrm{p})$ is obtained by integrating
over the variables
$z_1, \ldots, z_N$, leading to
\be \label{dist-Pzp}
Q_N(z_\mathrm{p}) = \frac{1}{Z} \, e^{-\beta U_\mathrm{p}(z_\mathrm{p})}
\left(\int_0^{z_\mathrm{p}} dz\, e^{-\beta U(z)}\right)^N.
\ee
Before going into more detailed calculations, we briefly discuss
two simple limiting cases of interest, namely $U(z)=0$ and $U(z)=U_\mathrm{p}(z)$.
The case $U(z)=0$ is the most standard equilibrium case, for which
\be
Q_N(z_\mathrm{p}) = \frac{1}{Z} z_\mathrm{p}^N \, e^{-\beta U_\mathrm{p}(z_\mathrm{p})},
\ee\label{ideal}
and Gaussian fluctuations are recovered in the large $N$ limit \cite{Wannier}.

In contrast, if the potential acting on the piston is the same
as that acting on the particles, one can express $P(z_\mathrm{p})$ in the
following form:
\be \label{eq-Pzp0}
Q_N(z_\mathrm{p}) = \frac{d}{dz_\mathrm{p}} G(z_\mathrm{p})^{N+1},
\ee
with
\be
G(z_\mathrm{p}) = \frac{\int_0^{z_\mathrm{p}} dz\,
e^{-\beta U(z)}}{\int_0^{\infty} dz\, e^{-\beta U(z)}}.
\ee
Volume fluctuations then exactly map onto an auxiliary problem of
extreme values, namely, the fluctuations of the maximal height of
$N+1$ independent particles (with no piston) with positions
$z_i'$ in a potential $U(z)$. The function $G(z_\mathrm{p})$ is simply
the probability that the position
of a particle subjected to the potential $U(z)$ is smaller than $z_\mathrm{p}$.
Hence, $G(z_\mathrm{p})^{N+1}$ is the probability that the positions of $N+1$
particles are less than $z_\mathrm{p}$, which is nothing but the cumulative
distribution of the maximum of the $N+1$ positions $z_i'$.
From Eq.~(\ref{eq-Pzp0}), it follows that $P(z_\mathrm{p})$ is the distribution
of $\max(z_1',\ldots,z_{N+1}')$, so that in the large $N$ limit,
fluctuations of $z_\mathrm{p}$ are described by standard extreme value
distributions.

In this paper, we are mostly concerned with the intermediate situation
where particles are submitted to a force derived from a potential,
but where this force is smaller than that acting on the piston.
A typical situation of this type is that of a system placed in a gravitational
field, as the piston generically has a mass larger than that of the particles.

Let us now compute the asymptotic volume distribution for general potentials
$U(z)$ and $U_\mathrm{p}(z)$. One way to tackle this issue could be to start
directly from Eq.~(\ref{dist-Pzp}).
Rather, as a short-cut, we take an alternative approach using
the results derived in \cite{BC06}. To this aim, we introduce the intervals
between the ordered positions of the particles in the following way.
For a given set of values $z_1,\ldots,z_N$ satisfying $0<z_i<z_\mathrm{p}$ for
all $i$, we introduce a permutation $\sigma$ of the integers $1,\ldots,N$
such that $z_{\sigma(1)} \le z_{\sigma(2)} \le \ldots \le z_{\sigma(N)}$,
and we define the space interval $h_i$ between particles through
\be \label{def-hi}
h_i = z_{\sigma(i)}-z_{\sigma(i-1)}, \qquad i=2,\ldots,N.
\ee
For convenience, we also introduce the variables $h_1$ and $h_{N+1}$,
\be \label{def-w1}
h_1=z_{\sigma(1)}, \qquad h_{N+1}=z_\mathrm{p}-z_{\sigma(N)}.
\ee
It is then straightforward to express $z_i$ as a function of the variables
$h_j$, namely
\be\label{heights}
z_i = \sum_{j=1}^{\sigma^{-1}(i)} h_j, \quad (i=1,\ldots, N), \qquad
z_\mathrm{p} = \sum_{j=1}^{N+1} h_j.
\ee
where $\sigma^{-1}$ is the inverse permutation of $\sigma$.
The system can then be described by the set
$(h_1,\ldots,h_{N+1})$, with $h_i \ge 0$ for all $i$, up to an arbitrary
permutation of the $N$ distinguishable particles. A given set
$(h_1,\ldots,h_{N+1})$ then corresponds to $N!$ configurations of the
particles (the position of the piston is fixed when the $h_i$'s are given). 
Summing over the corresponding $N!$ configurations $(z_1,\ldots,z_N)$
in Eq.~(\ref{eq-dist}), one obtains the equilibrium probability
distribution
\be \label{eq-PNh}
\tilde{P}_N(h_1,\ldots,h_{N+1}) = \frac{N!}{Z}\,
e^{-\beta U_\mathrm{p}(\sum_{j=1}^{N+1} h_j)}
\prod_{k=1}^N e^{-\beta U(\sum_{j=1}^k h_j)},
\ee
where we relabelled the factors in the product using $k=\sigma^{-1}(i)$.
Note that this last equation is also obtained in the case when particles
cannot cross each other. In this case, the 'no-crossing' constraint needs
to be taken into account from the outset in Eq.~(\ref{eq-dist}), and
Eq.~(\ref{eq-PNh}) is rather straightforwardly obtained, since there is
no need for reordering the positions of the particles.
Note also that the case of indistinguishable particles leads to a result
similar to that of distinguishable particles that cannot cross each other,
since in both cases it is not possible to generate a different
configuration through a permutation of the particles.

\begin{figure}[t]
\centering\includegraphics[width=7.5cm,clip]{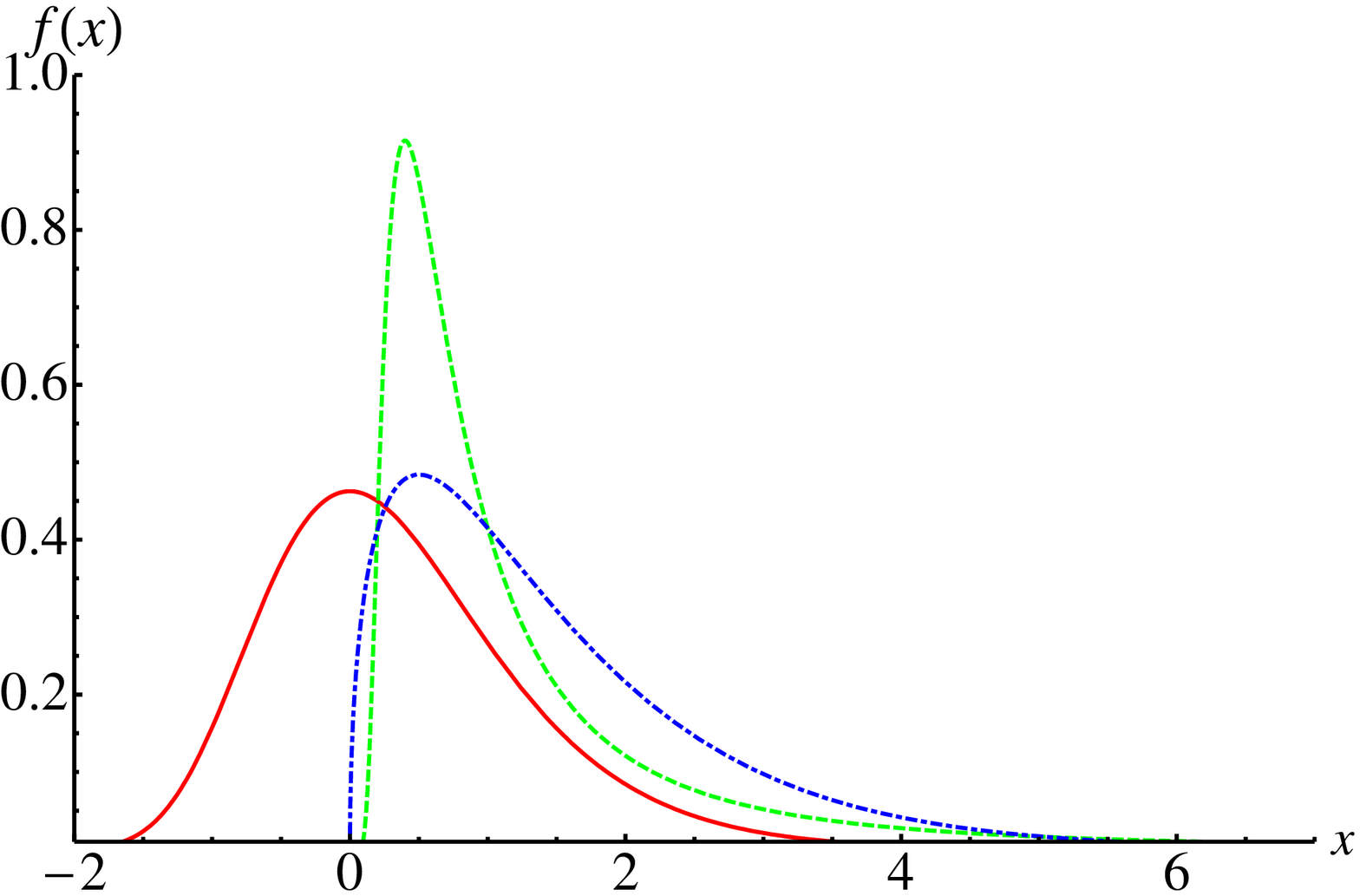}
\hfill
\centering\includegraphics[width=7.5cm,clip]{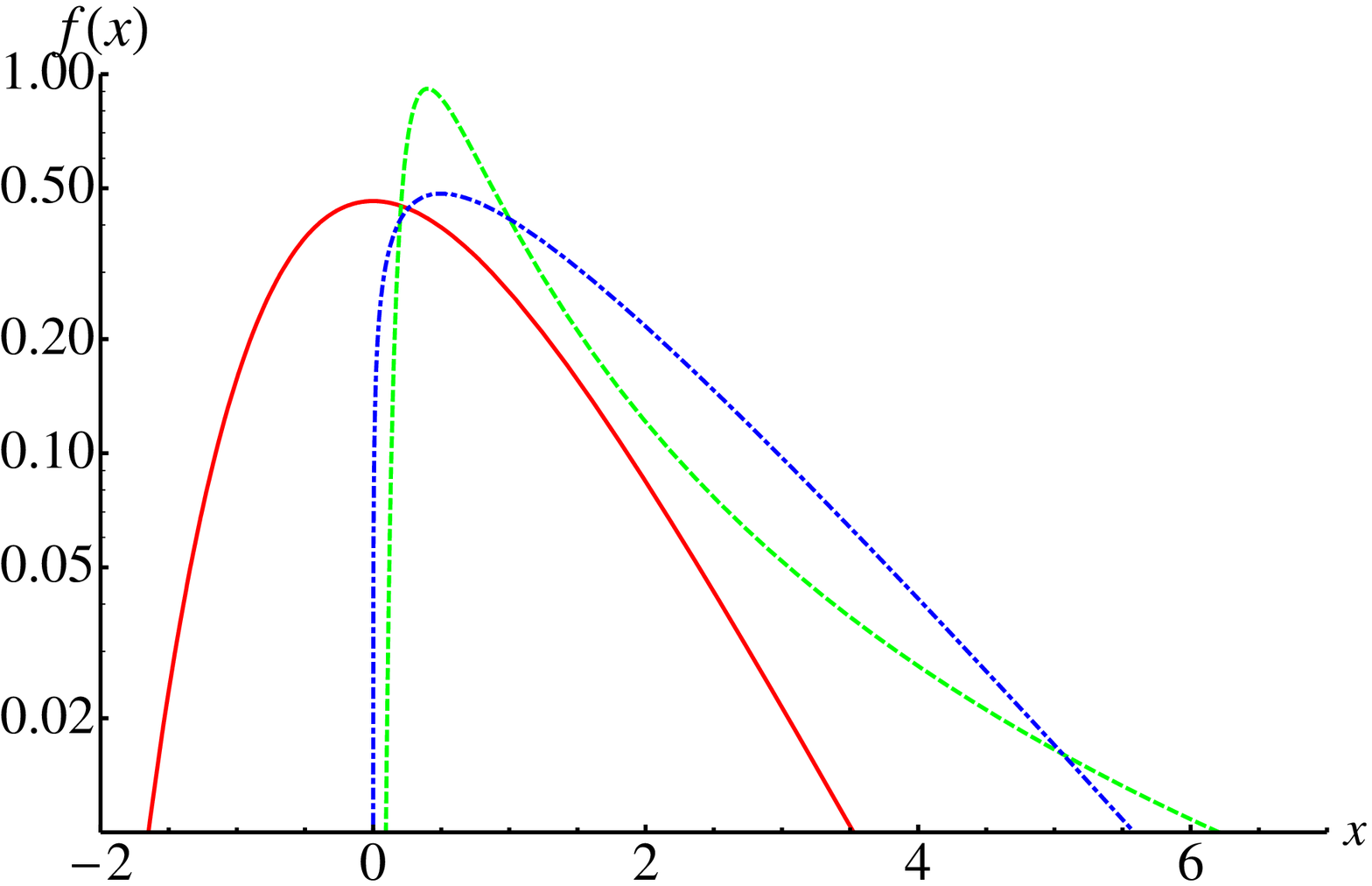}
\caption{\sl Examples of generalized extreme value distributions $f(x)$,
with $a=1.5$. Left: Gumbel distribution $f(x)=G_a(x)$ (full line),
Fr\'echet distribution $f(x)=F_{a,\mu}(x)$ with $\mu=1$ (dashed line)
and Weibull distribution $f(x)=W_{a,\mu}(x)$ with $\mu=1$ (dot-dashed).
Right: same distributions on a semi-logarithmic scale.}
\label{fig-EVS}
\end{figure}

The distribution $\tilde{P}_N(h_1,\ldots,h_{N+1})$ given in Eq.~(\ref{eq-PNh})
turns out to be quite similar to the joint distribution describing
the class of correlated random variables introduced in \cite{BC06}.
Up to slight notation changes \footnote{Starting from the distribution
$J_N(u_1,\ldots,u_N)$ defined in \cite{BC06},
we change $N$ into $N+1$, and reverse the order of index,
that is, we define the variables $h_i=u_{N+2-i}$.}, the latter reads:
\be \label{dist}
J_N(h_1,...,h_{N+1}) = K_N\;
\Omega\left[ F\left(\sum_{i=1}^{N+1}h_i\right)\right]
\prod_{i=1}^{N+1}P\left(\sum_{j=1}^{i} h_i\right).
\ee
$\Omega(F)$ is an arbitrary function of $F$, with $0<F<1$,
and the function $F(z)$ is defined as
\be \label{def-Fz}
F(z) = \int_z^{\infty} P(z') dz'.
\ee
The function $P(z)$ has the properties of a one-variable probability
distribution, namely it is a positive function such that
$\int_0^{\infty} P(z') dz'=1$.
In order to map the gas model onto Eq.~(\ref{dist}), we make the following
identification:
\bea
\label{id-Pz}
P(z) = \lambda\, e^{-\beta U(z)},\\
\Omega[F(z_\mathrm{p})] P(z_\mathrm{p}) = e^{-\beta U_\mathrm{p}(z_\mathrm{p})}
\label{id-Omega},
\eea
for all $z$, $z_\mathrm{p}>0$, and where $\lambda$ is a normalization factor. 
With this identification, Eq.~(\ref{eq-PNh}) can be rewritten as
\be
\tilde{P}_N(h_1,\ldots,h_{N+1}) = \frac{N!}{Z \lambda^N}\,
\Omega\left[F\left(\sum_{j=1}^{N+1} h_j\right)\right]
\prod_{k=1}^{N+1} P\left(\sum_{j=1}^k h_j\right),
\ee
which is precisely the same form as in Eq.~(\ref{dist}).
As $P(z)>0$ for all $z>0$, it results from Eq.~(\ref{def-Fz}) that $F(z)$ 
is a strictly decreasing function of $z>0$,
so that $y=F(z)$ can be inverted into $z=F^{-1}(y)$.
Accordingly, Eq.~(\ref{id-Omega}) can be reformulated as
\bea \nonumber
\Omega(y) &=& \frac{\exp[-\beta U_\mathrm{p}(F^{-1}(y))]}{P(F^{-1}(y))},\\
\label{def-Omega}
&=& \frac{1}{\lambda} \exp\left[\beta U(F^{-1}(y))-\beta U_\mathrm{p}(F^{-1}(y))\right],
\label{eq-Omega}
\eea
for all $y$, $0<y<1$. Eq.~(\ref{def-Omega}) actually gives a definition
of the function $\Omega(y)$.
The key result of \cite{BC06} is that if 
\be
\Omega(y) \sim \Omega_0\, y^{a-1}, \qquad y \to 0, \qquad (a>0),
\ee
the distribution of the sum
$z_\mathrm{p}=\sum_{i=1}^{N+1} h_i$ converges, up to a suitable rescaling,
to one of the generalized extreme value distributions with parameter $a$.
These distributions, illustrated in Fig.~\ref{fig-EVS},
belong to three different classes,
depending on the large $z$ behaviour of $P(z)$.
If $P(z)$ decays faster than any power law (typically, a power-law potential
$U(z)$), the generalized Gumbel distribution $G_a(x)$ is obtained, namely
\be \label{dist-Gumbel}
G_a(x) = C_\mathrm{g}\exp\left(-ay-ae^{-y}\right), \quad y=\theta(x+\nu),
\ee
where $\theta$ and $\nu$ are rescaling factors introduced to have zero mean
and unit variance, and $C_\mathrm{g}$ is a normalization factor
(see Appendix A).
In the limit $a \to \infty$, the generalized Gumbel distribution
converges to the Gaussian distribution.
If $P(z)$ has a power law tail $P(z)\sim z^{-1-\mu}$ when
$z \rightarrow \infty$, with $\mu>0$
(corresponding, in the present mapping, to
a logarithmic potential $U(z)$), one finds the generalized Fr\'echet
distribution,
\be \label{dist-Frechet}
F_{a,\mu}(x) = \frac{C_\mathrm{f}}{x^{1+a\mu}} \exp\left(-b_f x^{-\mu}\right),
\qquad x>0.
\ee
Finally, if $P(z)$ vanishes above a certain value $z_{\mathrm{max}}$
(say, there is a hard wall at $z_{\mathrm{max}}$), and behaves as a power-law
$P(z) \sim (z_{\mathrm{max}}-z)^{\mu-1}$, for $z\rightarrow z_{\mathrm{max}}$,
with $\mu>0$,
then the resulting distribution is of the generalized Weibull type,
\be \label{dist-Weibull}
W_{a,\mu}(x) = C_\mathrm{w} \, x^{a\mu-1} \exp\left(-b_w x^{\mu}\right),
\qquad x>0.
\ee
The parameters $b_f$ and $b_w$ are scale parameters, that can
be tuned to obtain any of the distributions with unit mean or with
unit variance (for the Fr\'echet distribution, this is only possible
if $\mu$ is large enough so that the mean or the variance are finite).

As a rather generic illustration, let us consider the case when the potentials
$U(z)$ and $U_\mathrm{p}(z_\mathrm{p})$ are given by
\be
U(z) = U_0\, z^{\alpha}, \qquad U_\mathrm{p}(z_\mathrm{p}) = U_0'\, z_\mathrm{p}^{\gamma},
\ee
with $\alpha$, $\gamma>0$.
If $\alpha=\gamma=1$, the particles and the piston are in a constant
external force field, like the (local) gravity field, in which case
$U_0=mg$ and $U_0'=Mg$.
If $\gamma=2$, the piston is for instance linked to the hard wall situated
at $z=0$ with a spring of stiffness $k=2U_0'$.

In order to determine the limit distribution of the volume fluctuations in
the large $N$ limit, the parameter $a$ characterizing the small $y$ behaviour
of $\Omega(y)$ should be evaluated.
Considering Eq.~(\ref{def-Omega}), one needs to compute first the
function $F^{-1}(y)$ in the small $y$ limit, which is deduced from
the large $z$ limit of $F(z)$. Given that
\be
P(z) = \lambda\, e^{-\beta U_0 z^{\alpha}}, \qquad
\lambda = \frac{\alpha(\beta U_0)^{1/\alpha}}{\Gamma\left(\frac{1}{\alpha}\right)},
\ee
with $\Gamma(t)=\int_0^{\infty} du\, u^{t-1}e^{-u}$ the Euler Gamma function,
one has in the large $z$ limit
\be
F(z) = \int_z^{\infty} \lambda\, e^{-\beta U_0 x^{\alpha}} dx
\approx \frac{(\beta U_0)^{\frac{1}{\alpha}-1}}{\Gamma\left(\frac{1}{\alpha}\right)\, z^{\alpha-1}}\, e^{-\beta U_0 z^{\alpha}}
\qquad (z \rightarrow \infty).
\ee
Inverting the relation $y=F(z)$ to get $z=F^{-1}(y)$, one finds
to leading order in the limit $y \rightarrow 0$
\be \label{eq-Fm1y}
F^{-1}(y) \approx \frac{1}{(\beta U_0)^{1/\alpha}}
\left(\ln\frac{1}{y}-\left(1-\frac{1}{\alpha}\right) \ln\ln\frac{1}{y}
-\ln\Gamma\left(\frac{1}{\alpha}\right)\right)^{1/\alpha}.
\ee
We now wish to compute $\Omega(y)$.
Putting Eq.~(\ref{eq-Fm1y}) into Eq.~(\ref{eq-Omega}), one needs
to distinguish between the cases $\alpha=\gamma$ and $\alpha \ne \gamma$.

If $\alpha=\gamma$, one gets
\be \label{Omega_id}
\Omega(y) \approx \frac{\Gamma\left(\frac{1}{\alpha}\right)^a}
{\alpha(\beta U_0)^{1/\alpha}} \, y^{a-1}
\left(\ln\frac{1}{y}\right)^{(a-1)(1-\frac{1}{\alpha})}
\qquad (y \rightarrow 0),
\ee
with $a=U_0'/U_0$. 
Therefore, $\Omega(y)$ behaves as a power law in the small $y$ limit,
up to logarithmic corrections which do not modify the asymptotic distribution
(see Appendix B). It follows that volume fluctuations are described
by a Gumbel distribution with parameter $a$ \cite{BC06}, which compares
the relative intensity of the compression force acting on the piston
and of the forces directly acting on the particles.
In the limit of a strong compression force $a \rightarrow \infty$,
volume fluctuations asymptotically become Gaussian, as the generalized
Gumbel distribution converges to the normal distribution in this limit.
In the opposit limit $a \ll 1$ where the external compression force is small
with respect to the forces acting on the particles, the generalized Gumbel
distribution converges to an exponential distribution (see Appendix A).

In the case where the confining potential for the gas and that for the piston
are of different functional form, $\alpha \ne \gamma$, 
one finds for $\Omega(y)$, dropping
logarithmic corrections as well as constants of order unity,
\be
\Omega(y) \sim \frac{1}{y}\, \exp\left[ -\beta U_0'\left(\frac{1}{\beta U_0}
\ln\frac{1}{y}\right)^{\gamma/\alpha}\right] \qquad (y \rightarrow 0).
\ee
Hence $\Omega(y)$ does not behave as a power law when $y \rightarrow 0$,
so that the results of \cite{BC06} do not apply.
We show in Appendix C that when $\gamma>\alpha$, the limit distribution is
a Gaussian law, while for $\gamma<\alpha$ the limit distribution
is exponential.
This result is consistent with the following intuitive argument.
When $\gamma>\alpha$, $y\Omega(y)$ decays faster
than any power law, and one expects this situation to be similar to the
large $a$ limit, for which a Gaussian distribution is recovered.
In contrast, when $\gamma<\alpha$, $y\Omega(y)$ decays slower
than any power law, which is expected to be similar to the limit
$a\rightarrow 0$, in which case one obtains an exponential distribution,
as shown in Appendix A.
The physical interpretation is that for $\gamma>\alpha$ the piston confines
the gas more strongly that the bulk confining potential,
so that the latter becomes irrelevant at large size,
leading to a regular confined gas with standard thermodynamic
properties. For $\gamma<\alpha$ the reverse is true; the gas is confined
by the bulk potential; the piston becomes irrelevant and its fluctuations
become those of a piece of flotsam driven by the fluctuations of the
confined gas below.

Note that we focused here on power law potentials, which correspond to
a quite natural class of potentials. However, one could also consider
logarithmic potentials $U(z)$ and $U_\mathrm{p}(z_\mathrm{p})$, which
would lead to Fr\'echet distributions given in Eq.~(\ref{dist-Frechet})
for the volume fluctuations.
Alternatively, if the potential diverges for a finite value $z_\mathrm{max}$,
(for instance by adding a rigid wall on top of the piston), the
asymptotic distribution of fluctuations would be of the Weibull type,
as described in Eq.~(\ref{dist-Weibull}).

\section{Discussion: relation with thermodynamics}
\label{sect-discussion}

In the present paper, we have shown that the volume fluctuations of
an ideal gas of classical and independent particles confined by
an algebraic  potential acting on both particles and piston 
along one dimension and by hard walls in the
perpendicular directions,
are described by generalized extreme value statistics.
In the simple case when the piston is identical to the particles,
the appearance of standard extreme value distributions is easily understood
from a direct mapping of volume fluctuations onto an extreme value problem
of independent and identically distributed (i.i.d.) random variables.

In a more general situation the piston can differ from the particles
in two ways; either the confining potential is of the same form but of
different amplitude, meaning that the restoring forces on the
particle and piston are different. Or, the functional form
is different. In the first situation non-Gaussian height fluctuations still
occur in the limit of large $N$ in the form of generalized extreme value
distributions parameterized by a real variable, $a=U_0'/U_0$.
If the restoring force on the piston becomes much larger than that
on the particles, the distribution crosses over to Gaussian.
Yet for fixed values of the forces, that is for fixed $a$,
the distributions are non-Gaussian for all system sizes and no crossover
occurs as a function of $N$.
For different functional forms, a crossover does occur as a function
of system size, either to Gaussian fluctuations if the piston is strongly
confined, or to one-body non-Gaussian statistics, with exponential
height fluctuations if the piston is less strongly confined. Our results are, in principle 
valid for a system of arbitrary scale perpendicular to the $z$ axis. However, 
non-Gaussian fluctuations should be observable for $a$ of order unity only,
which implies a piston of microscopic extent perpendicular to $z$,
ensuring that any experimental realization would be in the form of
a quasi-one dimensional sample (another possibility could be to
exert two different forces on the piston, that could be fine-tuned to
compensate almost exactly).

Non-Gaussian volume fluctuations have strong consequences for the
thermodynamics.
In previous work, non-Gaussian order parameter fluctuations
have been related to critical phenomena or to the fact that an ordered
phase is unstable in low dimensions \cite{BramwellPRE,Mermin}.
Analogous physics occurs for the models considered here: 
non-Gaussian volume fluctuations lead to singular thermodynamics.
To illustrate this we consider first the simplest case
where piston and particles are identical, of mass $m$ and confined
by a gravitational force, $mg$, so that $\alpha=\gamma=1$.
In this case the height difference
variables $h_j$ defined in (\ref{heights}), are independent
and exponentially distributed:
\be
P(h_j) = (N-j+1)\beta mg\, \exp[-(N-j+1)\beta mgh_j],
\ee
as seen from Eq.~(\ref{eq-PNh}). One then has a $1/f$-noise-like
spectrum \cite{Antal}, $\langle h_j\rangle  = k_BT/(N-j+1)mg$, giving directly
\be
\langle z_\mathrm{p} \rangle \approx z_0 \ln (N+1),
\qquad z_0=\frac{k_BT}{mg},
\ee
where $z_0$ is the characteristic length scale for the particles set
by the gravitational field. 
Defining volume $V=S z_p$ and external pressure, $P = mg/S$,
leads to an equation of state for the confined ideal gas
$P\langle V \rangle=k_BT \ln(N)$ in the large $N$ limit.
This singular non-extensive behavior signifies the crossover between
a system confined by a bulk potential (the gravitational field on the
particles) and an external constraint (the pressure imposed by the piston).
It is mathematically equivalent to the case of a thermally excited
one dimensional interface with long range interactions \cite{Antal}.

Let us now consider the case where the piston mass $M$ is different from
the mass $m$ of the particles.
In this case, the variables
$h_j$ are still exponentially distributed, but now with
\be
\langle h_j\rangle  = \frac{k_BT}{mg(N-j+a)}, \qquad
a = \frac{U_0'}{U_0} = \frac{M}{m},
\ee
similarly to the ``truncated $1/f$-noise'' considered in \cite{Bertin05,BC06}.
The average piston position $\langle z_\mathrm{p} \rangle$ is then given by
\be
\langle z_\mathrm{p} \rangle = \sum_{j=1}^{N+1} \langle h_j\rangle
= z_0 \sum_{j=1}^{N+1} \frac{1}{(N-j+a)}.
\ee 
For large $N$, one can approximate the sum by an integral, yielding
\be \label{eq-zp}
\langle z_\mathrm{p} \rangle \approx z_0 \ln \left(\frac{N}{a}+1 \right).
\ee
Introducing again the (external) pressure $P=Mg/S$, one finds the equation
of state
\be
P\langle V\rangle=a k_BT \ln \left(\frac{N}{a}+1 \right),
\ee
where now $a$ is a function of pressure.
Hence, again we find that non-Gaussian fluctuations are associated
with non-extensive thermodynamics.
The ratio $N/a=Nm/M$ compares the total mass $Nm$
of the particles to the mass $M$ of the piston
(the mass $M$ could also be an effective mass accounting for the constant
force $f_\mathrm{p}$, positive or negative, exerted by an operator:
$M=M_0-f_\mathrm{p}/g$).
If the piston becomes macroscopic with total mass, $M$, exceeding
that of the gas, then we move into the regime where $N/a \ll 1$.
In this regime, we recover the ideal gas equation of state,
$P\langle V\rangle=Nk_BT$, as well as Gaussian volume fluctuations.
In the opposite case where $N/a \gg 1$ (typically
if $a$ is finite and $N$ is large), one has non-Gaussian fluctuations
as described in Sect.~\ref{sect-discussion}, and the non-extensive equation
of state $P\langle V\rangle \approx a k_BT \ln(N/a)$.
Assuming that $a$ is large but finite, and scaling the number of particles,
$N$, with all other parameters held fixed,
one therefore begins in the extensive regime for small (but macroscopic)
$N$ and $V$. In this regime $\langle z_\mathrm{p} \rangle$ is less than
$z_0$ and the effect of the confining field on the particles is negligible.
On increasing $N$, one crosses over into the non-extensive regime
when this length scale in exceeded.
However, no crossover is observed in the statistics of fluctuations since
$a$ is large, and fluctuations are practically Gaussian even in the
non-extensive regime.
Taking $\alpha=\gamma\ne 1$ requires
more calculation but leads to essentially equivalent results.

A consequence of these results is that in the non-extensive regime
the volume fluctuations are abnormally small on the scale set by
the mean volume or the number of particles. Through the fluctuation dissipation
relation, this scale is given by 
\be
-\frac{\partial \langle V\rangle}{\partial P} = \frac{1}{k_BT}\left(\langle V^2\rangle - \langle V\rangle^2\right),
\ee
which, in the limit $N/a \gg 1$ is independent of $\langle V\rangle$.
Hence the isothermal compressibility,
$\kappa = -(1/\langle V\rangle)\partial \langle V\rangle /\partial P$,
a normally intensive measure of the fluctuations, varies as $1/\ln(N)$ and
scales to zero in the limit, $N\rightarrow \infty$.
Physically this result occurs
because the potential confining the particles within the bulk of the sample
suppresses the collective fluctuations present in the standard thermodynamic
regime. The logarithmic dependence is characteristic of a marginal situation
between the two regimes and is analogue to the marginal stability of an
ordered phase at the lower critical dimension, such as the 2D-XY model
\cite{BramwellPRE}, or one dimensional interface with long range
interactions \cite{Antal}.

The case where the piston is more confined than the particles
is best illustrated by removing the confining potential for the particles
in the above example while keeping that for the piston. One now trivially
finds the ideal gas equation of state, $P\langle V\rangle=Nk_BT$ and
regular thermodynamic fluctuations from Eq.~(\ref{ideal}).

\appendix

\section{Large and small $a$ limits of the generalized Gumbel distribution}

In this appendix, we wish to show that the generalized Gumbel distribution
$G_a(x)$ converges to the Gaussian distribution when $a \rightarrow \infty$,
and to the exponential distribution when $a \rightarrow 0$.
The distribution $G_a(x)$ is defined in Eq.~(\ref{dist-Gumbel}),
with $\theta$, $\nu$ and $C_g$ given by \cite{BC06}
\be \label{cst-Gumbel}
\theta^2 = \Psi'(a), \quad
\nu = \frac{1}{\theta}[\ln a - \Psi(a)], \quad
C_g = \frac{a^a \theta}{\Gamma(a)}.
\ee
The function $\Psi(a)$ is the digamma function defined as
\be
\Psi(a) \equiv \frac{d}{da} \ln\Gamma(a),
\ee
where $\Gamma(a)$ is the Euler Gamma function.

Let us start with the case $a \rightarrow \infty$, and determine the large
$a$ behaviour of the constants $\theta$ and $\nu$ given in (\ref{cst-Gumbel}).
Using Stirling's approximation for the Gamma function,
\be
\Gamma(a) \approx \sqrt{\frac{2\pi}{a}} a^a e^{-a} \qquad (a \to \infty),
\ee
one finds
\bea
\Psi(a) &=& \frac{d}{da} \ln\Gamma(a) \approx \ln a -\frac{1}{2a}\\
\theta^2 &=& \frac{1}{a} + \frac{1}{2a^2}
\eea
so that $\theta$ and $\nu$ are given for large $a$ by
\be
\theta = \frac{1}{\sqrt{a}}+\mathcal{O}(a^{-3/2}), \qquad
\nu =  \frac{1}{2\sqrt{a}}+\mathcal{O}(a^{-3/2}).
\ee
It follows that
\be
\theta(x+\nu) = \frac{x}{\sqrt{a}}+\frac{1}{2a}+\mathcal{O}(a^{-3/2}).
\ee
Hence for fixed $x$, $\theta(x+\nu)$ goes to zero when $a \rightarrow \infty$,
so that the term $\exp(-\theta(x+\nu))$ can be expanded to second order.
Inserting the different asymptotic expansion given above in the generalized
Gumbel distribution leads to, up to order $a^{-1/2}$ corrections in the
exponential,
\be
G_a(x) \approx \frac{e^a}{\sqrt{2\pi}} \exp\left[-\sqrt{a}x-\frac{1}{2}
-a\left(1-\frac{x}{\sqrt{a}}-\frac{1}{2a}+\frac{x^2}{2a}\right)\right],
\ee
yielding in the infinite $a$ limit the standard Gaussian distribution
\be
G_a(x) \rightarrow \frac{1}{\sqrt{2\pi}}\, e^{-x^2/2} \qquad
(a \rightarrow \infty).
\ee

In the opposite limit $a \rightarrow 0$, the Gamma function behaves
as $\Gamma(a) \approx 1/a$, so that $\Psi(a) \approx -1/a$ and
\be
\theta = \frac{1}{a}, \qquad \nu = a\ln a + 1.
\ee
The normalization factor $C_g$ reads
\be
C_g = \frac{a^a\, \theta}{\Gamma(a)} \approx e^{a\ln a} \rightarrow 1
\qquad (a \rightarrow 0).
\ee
Using $\theta(x+\nu)=(x+a\ln a+1)/a$, one finds for $G_a(x)$
\bea
G_a(x) &\approx& \exp\left[-(x+a\ln a+1)-a\,e^{-(x+a\ln a+1)/a}\right]\\
&\approx& e^{-(x+1)}\, \exp\left[-\,e^{-(x+1)/a}\right]
\qquad (a \rightarrow 0).
\eea
It is easily seen that the second factor in the r.h.s.~converges to
a Heaviside function for $a \rightarrow 0$,
\be
\exp\left[-\,e^{-(x+1)/a}\right] \rightarrow \Theta(x+1)
\qquad (a \rightarrow 0),
\ee
so that the generalized Gumbel distribution converges to the exponential
distribution with zero mean and unit variance
\be
G_a(x) \rightarrow e^{-(x+1)}\,\Theta(x+1) \qquad (a \rightarrow 0).
\ee

\section{Case $\alpha=\gamma$: Effect of logarithmic corrections}

The aim of this appendix is to show that the logarithmic corrections
appearing in Eq.~(\ref{Omega_id}) do not change the asymptotic distribution.
To that purpose, we follow closely the procedure of \cite{BC06},
and keep essentially the same notations.
Using Eq.~(\ref{dist}) and the results of \cite{BC06}, the distribution 
$Q_N(z_\mathrm{p})$ is given by
\be
Q_N(z_\mathrm{p}) = \frac{K_N}{N!} P(z_\mathrm{p}) \Omega(F(z_\mathrm{p}))
(1-F(z_\mathrm{p}))^N.
\ee
We assume for $\Omega(y)$ a power-law form with logarithmic corrections
when $y \rightarrow 0$, namely
\be \label{asympt-Omega}
\Omega(y) \approx \Omega_0 \, y^{a-1} \left(\ln\frac{1}{y}\right)^{\delta},
\qquad (y \rightarrow 0).
\ee
Let us define the value $z_N^*$ such that $F(z_N^*)=a/N$. In the limit
$N \rightarrow \infty$, $z_N^*$ diverges as $F(z_\mathrm{p})\rightarrow 0$ when
$z_\mathrm{p} \rightarrow \infty$. Introducing the auxiliary function
$g(z_\mathrm{p})=-\ln F(z_\mathrm{p})$, we perform the following
change of variables, to look at fluctuations around $z_N^*$:
\be
z_\mathrm{p} = z_N^* + \frac{v}{g'(z_N^*)}.
\ee
Expanding $g(z_\mathrm{p})$ in the neighbourhood of $z_N^*$ leads to 
\be
g(z_\mathrm{p})=g(z_N^*)+v+\epsilon_N(v), 
\ee
where 
\be
\lim_{N \rightarrow \infty} \epsilon_N(v)=0.
\ee
The distribution $\Phi_N(v)$ is obtained from $Q_N(z_\mathrm{p})$ as
\bea \nonumber
\Phi_N(v) &=& \frac{1}{g'(z_N^*)}\, Q_N(z_\mathrm{p})\\
&=& \frac{K_N}{N!} \frac{g'(z_\mathrm{p})}{g'(z_N^*)}
F(z_\mathrm{p}) \Omega(F(z_\mathrm{p})) (1-F(z_\mathrm{p}))^N
\eea
where we have used $P(z_\mathrm{p}) = g(z_\mathrm{p}) F(z_\mathrm{p})$.
In the large $N$ limit, we have, keeping $v$ fixed,
\bea
F(z_\mathrm{p}) \approx \frac{a}{N}e^{-v},\\
(1-F(z_\mathrm{p}))^N \rightarrow \exp \left(-a\, e^{-v} \right),\\
g'(z_\mathrm{p})/g'(z_N^*) \rightarrow 1.
\eea
Using Eq.~(\ref{asympt-Omega}), one obtains for large $N$,
as $F(z_\mathrm{p})\rightarrow 0$,
\bea
F(z_\mathrm{p}) \Omega(F(z_\mathrm{p})) &\approx& 
\Omega_0 \left(\frac{a}{N}\right)^a e^{-av}
\left(\ln\frac{N}{a}+v\right)^\delta\\
&\sim& \Omega_0 \left(\frac{a}{N}\right)^a e^{-av} (\ln N)^\delta,
\qquad (N \rightarrow \infty).
\eea
Altogether, one finds
\be \label{PhiNv-appB}
\Phi_N(v) \approx \frac{K_N}{N!}(\ln N)^\delta \, \Omega_0
\left(\frac{a}{N}\right)^a e^{-av} \exp \left(-a\, e^{-v} \right)
\ee
Let us now compute $K_N$, which is given by \cite{BC06}
\be
\frac{N!}{K_N} = \int_0^1 dy\, \Omega(y) (1-y)^N.
\ee
With the change of variable $v=u/N$, we get for large $N$
\be
\frac{N!}{K_N} = \frac{1}{N} \int_0^N du\, \Omega\left(\frac{u}{N}\right)
\left(1-\frac{u}{N}\right)^N
\approx \frac{1}{N} \int_0^N du\, \Omega\left(\frac{u}{N}\right) e^{-u}.
\ee
Using the small $y$ expansion of $\Omega(y)$, one has
\bea \nonumber
\frac{N!}{K_N} &=& \frac{1}{N} \int_0^N du\,\Omega_0
\left(\frac{u}{N}\right)^{a-1} \left(\ln \frac{N}{u}\right)^\delta e^{-u}\\
&\sim& \frac{\Omega_0}{N^a} (\ln N)^\delta \, \Gamma(a),
\qquad (N \rightarrow \infty).
\eea
Coming back to the distribution $\Phi_N(v)$, one finally obtains
from Eq.~(\ref{PhiNv-appB})
\be
\Phi_N(v) \rightarrow \Phi_{\infty}(v) =
\frac{a^a}{\Gamma(a)} \exp\left(-av-ae^{-v}\right),
\qquad (N \rightarrow \infty),
\ee
which is precisely the generalized Gumbel distribution.
In order to recover the standard expression $G_a(x)$ given in
Eq.~(\ref{dist-Gumbel}), one simply needs to introduce the normalized variable
$x$ through $v=\theta(x+\nu)$, with $\theta$ and $\nu$ defined in
Eq.~(\ref{cst-Gumbel}).

\section{Case $\alpha \neq \gamma$: Convergence of the volume distribution toward Gaussian and exponential laws}

In this appendix, we compute the asymptotic volume distribution in the
case where $\Omega(y)$ is given by
\be \label{Omega-appC}
\Omega(y) \sim \frac{1}{y}\exp\left[-b\left(\ln\frac{1}{y}\right)^{\eta}\right], \qquad (y \rightarrow 0),
\ee
with $\eta=\gamma/\alpha\ne 1$, $\eta>0$ (we refer to Sect.~\ref{sect-dist}
for notations), and $b>0$.
The derivation follows essentially the same steps as in Appendix B, but
also bears some similarities with that done in Appendix A.
The distribution $Q_N(z_\mathrm{p})$ is given by
\be
Q_N(z_\mathrm{p}) = \frac{K_N}{N!}\, g'(z_\mathrm{p})\, F(z_\mathrm{p})
\, \Omega(F(z_\mathrm{p})) \, (1-F(z_\mathrm{p}))^N
\ee
Using the form (\ref{Omega-appC}) of $\Omega(y)$ and recalling
that $F(z_\mathrm{p})=\exp[-g(z_\mathrm{p})]$, one finds
for large $z_\mathrm{p}$
\be
Q_N(z_\mathrm{p}) \approx \frac{K_N}{N!} \,g'(z_\mathrm{p})
\,\exp[-bg(z_\mathrm{p})^{\eta}]\, \left(1-e^{-g(z_\mathrm{p})}\right)^N
\ee
We now make the following change of variables:
\be
z_\mathrm{p} = z_N^*+\frac{v}{b\eta g'(z_N^*)g(z_N^*)^{\eta-1}}
\ee
where $z_N^*$ is defined by $g(z_N^*)=\ln N$. One then has for large $N$
\be
g(z_\mathrm{p}) \approx \ln N + \frac{v}{b\eta (\ln N)^{\eta-1}}, \qquad
g(z_\mathrm{p})^{\eta} \approx (\ln N)^{\eta}+\frac{v}{b}
\ee
The distribution $\Phi_N(v)$ reads
\be \label{PhiNv-appC}
\Phi_N(v) = \frac{K_N}{N!\,b\eta(\ln N)^{\eta-1}} \exp\left[-b(\ln N)^{\eta}-v
-\exp\left(-\frac{v}{b\eta(\ln N)^{\eta-1}}\right)\right]
\ee
In the case $\eta<1$, $(\ln N)^{1-\eta} \rightarrow \infty$ when
$N \rightarrow \infty$, so that
\be
\exp\left[-\exp\left(-\frac{v(\ln N)^{1-\eta}}{b\eta}\right)\right] \rightarrow \Theta(v),
\qquad (N \rightarrow \infty),
\ee
and $\Phi_N(v)$ converges to the exponential distribution
$\Phi_{\infty}(v)=e^{-v}\, \Theta(v)$ (the prefactor converges to $1$
by normalization of the distribution).

In the opposite case $\eta>1$, the above argument does not apply anymore.
Thus we start again from Eq.~(\ref{PhiNv-appC}), and make a saddle point
calculation. Let us introduce the function $\psi_N(v)$ such that
$\Phi_N(v) = \tilde{K}_N \exp(-\psi_N(v))$, namely
\be
\psi_N(v) = v+\exp\left(-\frac{v}{b\eta(\ln N)^{\eta-1}}\right)
\ee
with
\be
\tilde{K}_N = \frac{K_N}{N!\,b\eta(\ln N)^{\eta-1}} \,
\exp\left[-b(\ln N)^{\eta}\right]
\ee
We define $v^*$ through $\psi_N'(v^*)=0$, leading to
\be
\exp\left(-\frac{v^*}{b\eta(\ln N)^{\eta-1}}\right) = b\eta(\ln N)^{\eta-1}.
\ee
We then perform a change of variable
\be
v=v^*+x \sqrt{b\eta (\ln N)^{\eta-1}}.
\ee
For large $N$, we expand the term $\exp(-x/\sqrt{b\eta(\ln N)^{\eta-1}})$
appearing in $\psi_N(v)$ to second order, yielding
\be
\psi_N(v) = v^* + b\eta(\ln N)^{\eta-1} + \frac{x^2}{2}
+ \mathcal{O}\left( (\ln N)^{-(\eta-1)/2} \right)
\ee
The resulting distribution of $x$,
\be
\tilde{\Phi}_N(x) = \frac{1}{\sqrt{b\eta (\ln N)^{\eta-1}}}\,
\exp[-\psi_N(v)]
\ee
then converges to a Gaussian law when $N \rightarrow \infty$
(here again, the normalization of the distribution $\tilde{\Phi}_N(x)$
ensures that the prefactor converges to the correct limit).

\end{document}